\documentclass[letterpaper, 10 pt, conference]{ieeeconf}  

\IEEEoverridecommandlockouts                              

\usepackage{graphicx} 
\usepackage{amsmath,amsfonts,amssymb,bbold,dsfont}
\usepackage[font=footnotesize]{subfig}

\title{\LARGE \bf
A Multiple-Input Multiple-Output Cepstrum*
}

\author{Oliver Lauwers$^{1}$, Oscar Mauricio Agudelo$^{1}$ and Bart De Moor$^{1}$
\thanks{*Work financially supported by FWO-Flanders (EOS Project no 30468160 (SeLMA)), several PhD postdoc grants (FWO) and innovation mandate grant (VLAIO), EU H2020-SC1-2016-2017 (Grant 727721: MIDAS Meaningful Integration of Data, Analytics and Services), KU Leuven BOF C16/15/059 (nD Roots) \& C32/16/013. O. L. is a FWO SB fellow.}
\thanks{$^{1}$Stadius, Department of Electrical Engineering (ESAT),
	KU Leuven, 3000 Leuven, Belgium
	{\tt\footnotesize \{oliver.lauwers,mauricio.agudelo,}\newline{\tt\footnotesize bart.demoor\}@esat.kuleuven.be}}
}

\begin{document}

\maketitle
\thispagestyle{empty}
\pagestyle{empty}

\begin{abstract}

This paper extends the concept of scalar cepstrum coefficients from single-input single-output linear time invariant dynamical systems to multiple-input multiple-output models, making use of the Smith-McMillan form of the transfer function. These coefficients are interpreted in terms of poles and transmission zeros of the underlying dynamical system.

We present a method to compute the MIMO cepstrum based on input/output signal data for systems with square transfer function matrices (i.e. systems with as many inputs as outputs). This allows us to do a model-free analysis.

Two examples to illustrate these results are included: a simple MIMO system with 3 inputs and 3 outputs, of which the poles and zeros are known exactly, that allows us to directly verify the equivalences derived in the paper, and a case study on realistic data. This case study analyses data coming from a (model of) a non-isothermal continuous stirred tank reactor, which experiences linear fouling. We analyse normal and faulty operating behaviour, both with and without a controller present. We show that the cepstrum detects faulty behaviour, even when hidden by controller compensation. The code for the numerical analysis is available online.

\end{abstract}

\section{INTRODUCTION}

In this paper, we present an extension of the definition of the cepstrum to MIMO systems.
The main contributions of this paper are
\begin{itemize}
	\item the definition of the MIMO cepstrum,
	\item its interpretation in terms of poles and zeros,
	\item a computational scheme to estimate the MIMO cepstrum in the case of a system with as many inputs as outputs. This allows a model-free analysis of the underlying dynamics of input/output signals. We present a control theory case study on linear fouling in a non-isothermal continuous stirred tank reactor.
\end{itemize}
For the case where there are unequal numbers of inputs and outputs, a computational scheme to estimate the cepstrum from input and output signals is still lacking.

The cepstrum is a long-standing and versatile technique in signal processing, first discussed in \cite{bogert1963quefrency}. Originally, it was applied for the echo detection in seismic signals. The cepstrum has been used in a wide variety of applications, such as pitch detection in acoustic signals \cite{oppenheim1975digital}, analysis of mechanical problems \cite{randall2017history} and human activity recognition \cite{liu2016extreme}.

Applications are not only diagnostic, but also include parameter estimation \cite{oppenheim1975digital}, system identification and prediction \cite{DeCockThesis}, and assessing dynamical (dis)similarity between signals \cite{lauwers2017}. One of the main advantages of cepstral techniques is a rich theoretical framework, with an interpretation of the coefficients in terms of poles and zeros of the model.

This notion of the cepstrum of a signal was developed in the context of stochastic Linear Time Invariant (LTI) Single-Input Single-Output (SISO) dynamical systems. A major drawback is the absence of a notion of cepstral coefficients in the case of multiple inputs and multiple outputs (MIMO systems). While it is possible to calculate cepstra of each individual output and input, and create a cepstral coefficient matrix, it is not clear how this can be interpreted in terms of poles and zeros of the MIMO system as a whole.

The definition of the cepstrum to MIMO systems presented in this paper produces a scalar coefficient sequence, that reduces to the normal definition of the cepstrum in the SISO case, but preserves the interpretation in terms of poles and zeros of the system in the MIMO case.

This paper is structured as follows. Section \ref{sec:concepts} presents the concepts, notation and definitions used throughout the paper. Section \ref{sec:mimocepstrum} extends the cepstrum to the MIMO case, and its interpretation in terms of poles and zeros of the model. It also introduces an algorithm to compute the cepstrum in the case of square transfer matrices. Section \ref{sec:illustration} presents two numerical illustrations: a simple, fully-known synthetic model and a case study on a realistic dataset concerning linear fouling in a non-isothermal continuous stirred tank reactor, violating some of the assumptions made in introducing the cepstrum. The techniques presented in this paper will turn out to be quite robust and allow us to analyse this realistic scenario. Section \ref{sec:conclusions} will provide some general conclusions and possible paths for future work.

\section{CONCEPTS, NOTATION AND DEFINITIONS}
\label{sec:concepts}
In this section, we explain some concepts, notation and definitions that are used in the rest of the paper. In Subsection \ref{sub:siso}, we give a brief overview of the cepstrum in the traditional SISO framework and repeat its interpretation in terms of poles and zeros of the system. In Subsection \ref{sub:mimopolezero}, we provide an overview of the notion of poles and zeros of MIMO systems. Subsection \ref{sub:smithmcmillan} introduces the concept of Smith-McMillan forms of MIMO transfer matrices, which we employ in defining a MIMO cepstrum.

\subsection{SISO cepstrum}
\label{sub:siso}

A LTI SISO dynamical system can be represented as a \emph{state-space model}
\begin{equation}
\left\{
\begin{aligned}
x(k+1) &= Ax(k) + Bu(k)\\
y(k) &= Cx(k) + Du(k)
\label{eq:timedomain}
\end{aligned}
\right.
,
\end{equation}
where $k$ denotes (discrete) time, $x(k) \in \mathds{R}^n$ are the \emph{states} of the model, $u(k)$ and $y(k)$ \emph{input} and \emph{output} sequences respectively and $A$, $B$, $C$ and $D$ \emph{system matrices} of appropriate dimensions. We assume the model to be \emph{minimal} (i.e., \emph{observable} and \emph{controllable}).

Using the \emph{$z$-transform} from \cite{kailath1980linear} with $x(0) = 0$, this can be written as
\begin{equation}
Y(z) = H(z)U(z),
\label{eq:frequencydomain}
\end{equation}
where $U(z)$ and $Y(z)$ are the \emph{z-transform} of input and output respectively, and $H(z)$, the \emph{transfer function} of the system, which is the z-transform of the \emph{impulse response} of the system.

The transfer function is a rational function of $z$, with both numerator and denominator polynomials. We can write
\begin{equation}
H(z) = g\frac{b(z)}{a(z)},
\label{eq:transferfunction}
\end{equation}
where $g \in \mathds{R}$ is the constant \emph{gain} of the system, and $a(z)$ and $b(z)$ are monic polynomials of degree $n$, the roots of which are respectively \emph{poles} (denoted by $\alpha_i, \text{ } i \in \{1,2,\ldots,n\}$) and \emph{zeros} (denoted by $\beta_i, \text{ } i \in \{1,2,\ldots,n\}$) of the system. For simplicity we assume the system to be \emph{stable} (i.e. $|\alpha_i| < 1, \text{ } \forall i$), \emph{causally invertible} (i.e. $n$ poles and $n$ zeros)\footnote{Note that we can assume this without loss of generality. For the purpose of this paper, we can always add zero's at $z=0$ without changing results.}, \emph{minimum-phase} (i.e. $|\beta_i| < 1, \text{ } \forall i$) and \emph{minimal} (i.e. $\alpha_i \neq \beta_j, \text{ } \forall i, \text{ } \forall j$). To keep notation simple, we assume that all poles and zeros are \emph{simple} (i.e., they have multiplicity 1) throughout this text. Extensions to multiple poles and zeros are possible but would burden notation.

The transfer function leads to the notion of \emph{power spectral density}, defined as \cite{kailath1980linear}
\begin{equation}
\hspace{-0.195em}\Phi_H(\textnormal{e}^{i\omega}) = H(\textnormal{e}^{i\omega})\overline{H(\textnormal{e}^{i\omega})} = \left|H(\textnormal{e}^{i\omega})\right|^2 = \left|g\frac{b(\textnormal{e}^{i\omega})}{a(\textnormal{e}^{i\omega})}\right|^2.
\label{eq:powerspectrum}
\end{equation}
Here, the subscript $\cdot_H$ denotes that it is the power spectral density of the transfer function. Similar notation for input and output leads to $\Phi_U$ and $\Phi_Y$. Here, $U(\textnormal{e}^{i\omega})$ and $Y(\textnormal{e}^{i\omega})$ are estimated empirically using \emph{Welch's method} \cite{lauwers2017,welch1967use}. The overbar denotes the complex conjugate, $i$ the imaginary unit and $\omega$ is the angular frequency. 

The \emph{(power) cepstrum}\footnote{The terminology \emph{power} cepstrum comes from the fact that it is derived from the \emph{power} spectrum of the signal. A similar concept, based on the z-transform itself, is known as the \emph{complex} cepstrum. In this paper, we only discuss the power cepstrum. We will use the terms power cepstrum and cepstrum interchangeably.} of a system is then defined as
\begin{equation}
c_H(k) = \mathcal{F}^{-1}(\log\Phi_H(\textnormal{e}^{i\omega})),
\label{eq:definitionpowercepstrum}
\end{equation}
where $\mathcal{F}^{-1}$ is the \emph{inverse Fourier transform}. Similar definitions hold for the cepstra of input and output, $c_U$ and $c_Y$.

The rationale behind the use of the cepstrum in signal processing comes from \emph{homomorphic signal processing} \cite{oppenheim1975digital}. A look at Equation \eqref{eq:definitionpowercepstrum} shows us that the cepstrum takes convolutions in the time domain to a multiplication in the power spectral domain and then to additions via the logarithm. To return to (a transformed version of) the time domain, the inverse Fourier transform is applied. The original convolutional structure is thus equivalent to an additive one. In other words, a convolution of two signals in the time domain, $u = u_1 \ast u_2$, will result in an addition of their cepstra, $c_U = c_{U_1} + c_{U_2}$.

These cepstrum coefficients can be interpreted in terms of poles and zeros of the underlying system. In particular, for a system with transfer function
\begin{equation}
H(z) = g\frac{b(z)}{a(z)} = g\frac{\prod_{i=1}^n(1-\beta_iz^{-1})}{\prod_{i=1}^n(1-\alpha_iz^{-1})},
\label{eq:transferfunctionpoleszeros}
\end{equation}
we can show (see the Appendix for a derivation) that
\begin{equation}
\begin{aligned}
c_H(k) &= \sum_{j=1}^{n}\frac{\alpha_j^{|k|}}{|k|}  
- \sum_{j=1}^{n}\frac{\beta_j^{|k|}}{|k|}
\hspace{7pt} \forall k \neq 0,\\
c_H(0) &= \log(g^2).
\end{aligned}
\label{eq:cepstrumpoleszeros}
\end{equation}

These expressions link the power cepstrum to the poles and zeros of the underlying dynamics. It is this connection that makes the cepstrum a powerful signal processing technique. E.g., this property allows the cepstrum to be employed to define a similarity measure that takes into account underlying dynamics of signals \cite{lauwers2017,de2000subspace2}. It is therefore a logical conclusion to demand that any extension of the cepstrum to the MIMO case should retain this interpretability.

In what follows, we will define an extension that is connected in a similar way to poles and zeros of the MIMO system.

\subsection{MIMO poles and zeros}
\label{sub:mimopolezero}

For a system with $l$ inputs and $m$ outputs, we have $u(k) \in \mathds{R}^l$ and $y(k) \in \mathds{R}^m$ in the state-space model \eqref{eq:timedomain}. The transfer function \eqref{eq:transferfunctionpoleszeros} is no longer a rational function, but a $m\times l$ matrix, the elements of which are rational functions. 

For the purpose of this paper, we take the following definition of poles and zeros:
\begin{description}
	\item[\textbf{Zeros}] are transmission zeros, i.e., complex numbers for which the transfer matrix,
	\begin{equation}
	H(z) = D + C(z\mathds{1}-A)^{-1}B,
	\end{equation}
	drops below its \emph{normal rank, $r$} (i.e. $r$ is the rank of the matrix $H(z)$ for almost all points in $\mathds{C}$).
	\item[\textbf{Poles}] are the eigenvalues of the $A$ matrix in Equation \eqref{eq:timedomain}.
\end{description}
Extensive discussion of these properties can be found in many works, e.g. \cite{kailath1980linear}, and we will not repeat these here.
The goal in this paper is to define a MIMO cepstrum that can be interpreted in terms of these definitions of poles and zeros in an analogous way to Equation \eqref{eq:cepstrumpoleszeros}. We introduce the Smith-McMillanform, making the connection between the transfer matrix and its poles and zeros more explicit.
\subsection{Smith-McMillan form}
\label{sub:smithmcmillan}

A transfer matrix is a matrix of rational transfer functions. In this section, we explain the \emph{Smith-McMillan form} of a rational matrix. This form of the transfer function is a pseudo-diagonal matrix, with non-zero diagonal elements consisting of polynomials with roots equal to some of the poles and zeros (as defined in Subsection \ref{sub:mimopolezero}). For the definition in this Subsection, we rely heavily on \cite{kailath1980linear}.

Pseudo-diagonalising these matrices is done by applying \emph{elementary operations} on a rational matrix, which are
\begin{itemize}
	\item multiplication of a row/column by a constant,
	\item switching positions of two rows/columns,
	\item addition of a polynomial multiple of one row/column to another.
\end{itemize}
These elementary operations can be represented as matrices, which multiply a rational matrix from the left for row operations, and from the right for column operations.

Combining the corresponding row operations into a \emph{unimodular} (i.e. of constant determinant) polynomial matrix $V_1(z)$, and the corresponding column operations into a unimodular polynomial matrix $V_2(z)$, we can write for a transfer matrix $H(z)$
\begin{equation}
V_1(z)H(z)V_2(z) = M(z),
\label{eq:smithmcmillan}
\end{equation}
with
\begin{equation}
M(z) = \begin{pmatrix}
\text{diag}\left\{g_i\frac{b_i(z)}{a_i(z)}\right\} & 0\\
0 & 0
\end{pmatrix},
\label{eq:diagonal}
\end{equation}
where $a_{i+1}(z) | a_i(z)$ (i.e. $a_{i+1}(z)$ exactly divides $a_{i}(z)$), $b_{i}(z) | b_{i+1}(z)$ and $i\in \{1,2,\ldots,r\}$, with $r$ the normal rank of $H(z)$. The $g_i$ are constant gains.

We denote
\begin{equation}
b(z) = \prod_{i=1}^{r} b_i(z), \hspace{2em} a(z) = \prod_{i=1}^{r} a_i(z),
\label{eq:zeropolepolynomial}
\end{equation}
the \emph{zero} and \emph{pole polynomial} respectively, i.e. the solutions of $b(z) = 0$ and $a(z) = 0$, equal to the zeros and poles of the transfer matrix $H(z)$. In fact, the Smith-McMillan form provides an alternative yet equivalent way of defining poles and zeros of transfer matrices. We do not prove this, but a detailed discussion can be found in \cite{kailath1980linear}.

\section{MIMO cepstrum}
\label{sec:mimocepstrum}
For the MIMO case, the assumptions on the transfer matrix are the same as in the SISO case, presented directly after Equation \eqref{eq:transferfunction}, with one notable exception: as the transfer matrix is of size $m \times l$, and therefore not necessarily square, we cannot assume invertibility of $H(z)$. We replace it by an assumption on the normal rank of the transfer matrix:
\begin{equation}
r = \min\{m,l\}.
\label{eq:normalrank}
\end{equation}
Based on the size of $H(z)$ (i.e. the amount of inputs, $l$ relative to the amount of outputs, $m$), this assumption is equivalent to:
\begin{description}
	\item[$\mathbf{m>l}$] we assume \emph{left-invertibility}, i.e., there exists an $l\times m$ matrix $L(z)$ such that $L(z)H(z) = \mathds{1}_l$, where $\mathds{1}_l$ is the $l\times l$ identitiy matrix; this matrix $L(z)$ is the \emph{left-inverse} of $H(z)$,
	\item[$\mathbf{m<l}$] we assume \emph{right-invertibility}, i.e., there exists an $m\times l$ matrix $R(z)$ such that $H(z)R(z) = \mathds{1}_m$, where $\mathds{1}_m$ is the $m\times m$ identitiy matrix; this matrix $R(z)$ is the \emph{right-inverse} of $H(z)$,
	\item[$\mathbf{m=l}$] we assume the matrix to be invertible.
\end{description}

\subsection{Definition of the MIMO cepstrum}
\label{sub:mimodefinition}

With the assumptions made above, we extend the cepstrum to the MIMO case, for a transfer matrix $H(z)$ with Smith-Mcmillan form $M(z)$ and $m<l$, as
\begin{equation}
c_H(k) = \mathcal{F}^{-1}\left(\log\det \left( M(\textnormal{e}^{i\omega})\overline{M(\textnormal{e}^{i\omega})}^\intercal\right)\right),
\label{eq:definitionmimopowercepstrum}
\end{equation}
with $\cdot^\intercal$ denoting the matrix transpose. For the case where $m>l$ the definition is the same, but with the position of the transpose switched. For $m=l$, these are equivalent.
 
In the next Subsection, we will show how to interpret this extension of the cepstrum in terms of poles and zeros of the transfer matrix.

\subsection{Interpretation in terms of poles and zeros}
\label{sub:mimocepstrumpoleszeros}

In this section, we work with the assumption that the transfer matrix is right-invertible (i.e., $m<l$). The other cases are completely analogous, with transposes switching places, and we will not repeat the derivation.

From Equation \eqref{eq:diagonal}, and the assumption that $H(z)$ (and therefore $M(z)$) has normal rank $r = \min\{m,l\} = m$, we have (with $i \in \{1,2,\ldots,r\}$)
\begin{equation}
M(\textnormal{e}^{i\omega})\overline{M(\textnormal{e}^{i\omega})}^\intercal = diag\left\{\left|g_i\frac{b_i(\textnormal{e}^{i\omega})}{a_i(\textnormal{e}^{i\omega})}\right|^2\right\}.
\end{equation}
Taking the determinant, and applying Equation \eqref{eq:zeropolepolynomial}, we find
\begin{equation}
\det\left(M(\textnormal{e}^{i\omega})\overline{M(\textnormal{e}^{i\omega})}^\intercal\right) = \left|g\frac{b(\textnormal{e}^{i\omega})}{a(\textnormal{e}^{i\omega})}\right|^2,
\end{equation}
with $g = \prod_i{g_i}$ the product of the individual gains. This ratio of the zero and pole polynomial, however, is equivalent to the SISO case, where these polynomials are directly encapsulated in the transfer function \eqref{eq:transferfunctionpoleszeros}. The MIMO problem can now be solved in exactly the same way, following the derivation in the Appendix. We again get the result
\begin{equation}
\begin{aligned}
c_H(k) &= \sum_{j=1}^{n}\frac{\alpha_j^{|k|}}{|k|}  
- \sum_{j=1}^{n}\frac{\beta_j^{|k|}}{|k|}
\hspace{7pt} \forall k \neq 0.\\
c_H(0) &= \log(g^2).
\end{aligned}
\label{eq:cepstrumpoleszerosmimo}
\end{equation}
This gives an interpretation of the MIMO cepstrum in terms of poles and zeros of the underlying model.

The power cepstrum can now be readily derived whenever the Smith-McMillan form \eqref{eq:smithmcmillan} is available. In general, it is not straightforward to compute it given only input and output signals. When $m\neq l$, we have no way of computing the cepstrum without explicitly estimating a model and deriving the Smith-McMillan form. For the case where $m = l$, we present a way to do so in the next Subsection.

\subsection{Computation when $m=l$}
\label{sub:squarecomputation}
When there are as many inputs as outputs, we will prove that definition \eqref{eq:definitionmimopowercepstrum} is equivalent (except for $k=0$) to
\begin{equation}
c_{H,\text{comp}}(k) = \mathcal{F}^{-1}(\log\det\Phi_H(\textnormal{e}^{i\omega})),
\label{eq:computationalcepstrum}
\end{equation}
with $\Phi_h$ the transfer matrix power spectrum, defined as
\begin{equation}
\Phi_H(\textnormal{e}^{i\omega}) = H(\textnormal{e}^{i\omega})\overline{H(\textnormal{e}^{i\omega})}^\intercal.
\end{equation}
We can calculate the determinant of the power spectrum, using the Smith-McMillan form \eqref{eq:smithmcmillan}, as
\begin{equation}
\det\Phi_H = \det \left(V_1^{-1}MV_2^{-1}\overline{V_1^{-1}M V_2^{-1}}^\intercal\right).
\end{equation}
Here, $V_1^{-1}$ and $V_2^{-1}$ are the inverse of the unimodular matrices in Equation \eqref{eq:smithmcmillan}, and therefore themselves unimodular. We dropped the variables to make notation easier, but we understand all the matrices involved to be evaluated on the unit circle, e.g., $\Phi_H = \Phi_H(\textnormal{e}^{i\omega})$. Working out the transpose, and using the fact that the determinant is a multiplicative map\footnote{This is the troublesome step when $m\neq l$ and there will not necessarily be a straightforward equivalence between $\det\Phi_H$ and $\det\left(M\overline{M}^\intercal\right)$. A generalization of the multiplicative property, the Binet-Cauchy theorem \cite{knill2014cauchy,vishwanathan2004binet}, may offer a solution, which we will not explore further here.} for square matrices (i.e, for square matrices $A$ and $B$, $\det(AB) = \det(A)\det(B)$), we can write 
\begin{equation}
\det\Phi_H = \det\left(V_1^{-1}\overline{V_1^{-1}}^\intercal\right)\det\left(M\overline{M}^\intercal\right)\det\left(V_2^{-1}\overline{V_2^{-1}}^\intercal\right). 
\end{equation}
Unimodular matrices are matrices that have, by definition, a constant determinant. Denote the determinant of $V_1^{-1}$ as $c_{V_1}$ and that of $V_2^{-1}$ as $c_{V_2}$, which leads to
\begin{equation}
\det\Phi_H = |c_{V_1}|^2|c_{V_2}|^2\det\left(M\overline{M}^\intercal\right). 
\end{equation}
Using this, and comparing Equation \eqref{eq:computationalcepstrum} and \eqref{eq:definitionmimopowercepstrum},
\begin{equation}
c_{H,\text{comp}}(k) = c_H(k) + \mathcal{F}^{-1}\log\left(|c_{V_1}|^2|c_{V_2}|^2\right).
\end{equation}
Following the derivation in the Appendix, we see that this leads to
\begin{equation}
\begin{aligned}
c_{H,\text{comp}}(k) &= c_H(k) \hspace{8em} \forall k\neq 0,\\
c_{H,\text{comp}}(0) &= c_H(0) + \log\left(|c_{V_1}|^2|c_{V_2}|^2\right).
\end{aligned}
\end{equation}

While it is true that we cannot, in general, know the size of the error of our estimation on $c_H(0)$, in practical applications of the cepstrum (see for example the distances defined in \cite{martin2000metric,de2000subspace2,lauwers2017}), this coefficient is often not important, as it contains only information on the gain of the system.

One last step, to be able to compute the MIMO cepstrum based on input/output data of a system, is to estimate $\log\det\Phi_H$. We know $\Phi_Y = \Phi_H\Phi_U$, and easily see that, for square systems,
\begin{equation}
\log\det\Phi_H = \log\det\Phi_Y-\log\det\Phi_U.
\end{equation}
Equivalently, since the inverse Fourier transform is a linear operator, we can calculate the cepstra of input and output with Equation \eqref{eq:computationalcepstrum} and write 
\begin{equation}
c_{H,\textnormal{comp}}(k) = c_{Y,\textnormal{comp}}(k) - c_{U,\textnormal{comp}}(k), \hspace{1em} \forall k\neq 0.
\label{eq:additive}
\end{equation}

Since $u(k)$ and $y(k)$, the input and output signals, are given, we have found a data-driven, model-free way to compute the MIMO cepstrum for systems with $m=l$.

In the next Section, we will give some numerical illustrations and applications.
\section{ILLUSTRATION AND APPLICATION}
\label{sec:illustration}
In this Section, we first give a numerical illustration on synthetic data coming from a simple model. This will provide insight in the techniques presented in this paper. Afterwards, we show numerical results for a case study on faults in a Non-isothermal Continuous Stirred Tank Reactor. This is a realistic dataset and violates some of the assumptions made in this text. The cepstrum turns out to be quite robust. The code used to analyse these models is available online\footnote{https://github.com/Olauwers/MIMOcepstrum}.

\subsection{Numerical Illustration}
\label{sub:numericalillustration}
As a numerical illustration, we implement (in MATLAB) a simple MIMO system with 3 inputs and 3 outputs. The poles, zeros and gains of the individual entries are chosen randomly, but it is made sure that the transmission zeros of the MIMO system are minimum-phase. The seed of the random number generator was set to default, for reproducibility. The system is generated as a state-space model, which behaves better numerically in the MIMO case. The transmission zeros of the MIMO system are $\beta_i = \{-0.9681, 0.4419, {0.0916 \pm 0.1453i}\}$ and the poles are $\alpha_i = \{{0.1786 \pm 0.3300i}, {-0.2769 \pm 0.1793i}, 0.0634\}$.

\begin{figure}[t]
	\centering
	\hspace{-2em}
	\includegraphics[width=0.9\linewidth]{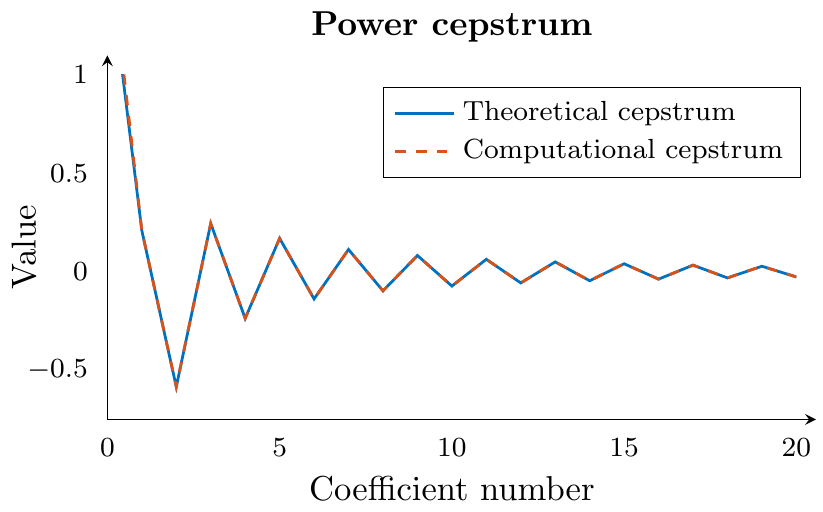}
	\caption{
		The theoretical cepstrum from Equation \eqref{eq:cepstrumpoleszerosmimo} and the estimate presented in Subsection \ref{sub:squarecomputation} for the synthetic data from Subsection \ref{sub:numericalillustration}. We see that the computational cepstrum is a very good estimate of the exact one. The differences (for $k\neq 0$) are of order $10^{-4}$.}
	\label{fig:numericalillustration}
\end{figure}

We generate a white noise input of length $2^{16}$, with random gains for every individual input channel. The computation in Subsection \ref{sub:squarecomputation} is used as an estimate of the cepstrum, and compared with the exact cepstrum in Equation \eqref{eq:cepstrumpoleszerosmimo}. Results are shown in Fig. \ref{fig:numericalillustration}. The computational cepstrum is indeed a very good estimate of the exact one. 

The white noise data results in a power cepstrum that is non-zero for $c_H(0)$, but vanishes everywhere else, analogous to the SISO case \cite{DeCockThesis}. This was to be expected, but provides an extra argument that the definition presented in this text is a natural generalization.

\subsection{Case study: Fouling In Non-isothermal Continuous Stirred Tank Reactor}

Continuous Stirred Tank Reactors (CSTRs) are one of the most important and fundamental units in chemical industry. They are characterized by highly nonlinear dynamics and they pose a challenging problem for fault prognosis and early fault detection algorithms. The CSTR considered in this section is a non-isothermal reactor where a single, first order, irreversible, and exothermic reaction takes place ($ A \rightarrow B$).

The model of the reactor consists of the following nonlinear  ordinary differential equations:
\small\begin{align}
	&\frac{dC_A}{dt}  = \frac{q}{V}\left( C_{A_{f}} - C_A \right) - C_A k_0e^{-\frac{E}{RT}} + v_1 \label{eq2}  \\
	&\frac{dT}{dt}  =  \frac{q}{V}\left( T_f - T \right) -\frac{\Delta H}{\rho C_p}C_Ak_0 e^{-\frac{E}{RT}} + \frac{UA}{V\rho C_p} \left( T_j - T  \right) + v_2 \nonumber
\end{align}\normalsize
\noindent with $C_A$ and $T$ the reactant concentration and temperature respectively inside the reactor, $T_j$ the jacket temperature, $q$ the flow rate of the feed flow, $T_f$ the temperature of the feed flow, $C_{A_f}$ the concentration of the reactant in the feed flow, and $v_1$, $v_2$ independent system noise processes, with $v_i \sim N(0,\sigma_{v_i}^2 = 0.01)$. Parameters of the model and their numerical values (taken from  \cite{cstr_paper}) are: $V = 100$ L (volume of the mix), $k_0 = e^{13.4} \ \rm{min}^{-1}$  (kinetic constant), $E/R = 5360 \ \rm{K}$ ($E$ is the activation energy and $R$ is the ideal gas constant), $(-\Delta H) = 17835.821 \ \rm{J}/\rm{mol}$ (heat of the reaction), $\rho = 1000 \ \rm{g}/\rm{L}$ (density), $C_p = 0.239 \ \rm{J} /\rm{g} / \rm{K}$ (specific heat), and $UA = 11950 \ \rm{J} / \rm{min} / \rm{K}$ ($U$ is the overall heat transfer coefficient, $A$ is the area of the heat exchange between reactor wall and jacket) in ideal conditions (no fouling). Under normal conditions, the operating point of the reactor is given by $C_{A}^* = 0.2  \ \rm{mol}/ \rm{L}$, $T^* = 446  \ \rm{K}$, $q^* = 100 \ \rm{L}/ \rm{min}$, $T_j^* = 419 \ \rm{K}$, $T_f^* = 400 \ \rm{K}$ and $C_{A_f}^* = 1 \ \rm{mol}/\rm{L}$. 

A control system consisting of two PID controllers and a decoupler keeps $C_A$ and $T$ around their nominal values $C_A^*$ and $T^*$, by manipulating the jacket temperature $T_j$ and the flow rate of the feed flow $q$. The control law in the Laplace domain is as follows \cite{cstr_paper}:
\begin{align}
\hspace{-1.5em}\left[ \begin{array}{c} q(s) \\ T_j(s) \end{array} \right] = & \left[\begin{array}{c c} 5 & 1 \\ 1 & 2 \end{array} \right] \left[
\begin{array}{c}
(K_p + K_ds + K_i/s) E_{C}(s) \\
(K_p + K_ds + K_i/s) E_T(s)
\end{array}
\right]
\end{align}
with $E_{C}(s) = C_A(s) - C_A^*(s)$, $E_T(s) = T(s) - T^*(s)$, $K_p = 1$, $K_d = 0.1$ and $K_i = 10$.

Fouling,  the accumulation of unwanted material on a heat transfer surface that increases its thermal resistance, is one of the most serious issues in heat transfer equipment. We distinguish two types of fouling: asymptotic and linear. In asymptotic fouling the resistance to heat transfer increases fast when the operation starts and becomes asymptotic to a steady state value at the end. In linear fouling the resistance increases linearly during the entire process operation. Here, we consider the second type, linear fouling. The overall heat transfer coefficient $U$  multiplied by the heat exchange area $A$ in (\ref{eq2}) is given by the equation ($t$ in minutes)
\begin{equation}
U(t)A=
\begin{cases}
11950,  & t \le 5000 \\
11950 - 0.8365 (t-5000), & t>5000
\end{cases}.
\end{equation}

Two datasets of 10000 points have been generated (sampling time  = 1 min), one when the controller is active and one when the controller has been switched off. Only the controlled ($C_A$ and $T$) and manipulated variables ($q$ and $T_j$) are measured, which are contaminated with measurement noise: $\eta_{\rm C_A} \sim N(0,\sigma_{C_A}^2 = 10^{-5})$, $\eta_{\rm T} \sim N(0,\sigma_{T}^2 = 0.005)$, $\eta_q \sim N(0,\sigma_{q}^2 = 10^{-6})$, and $\eta_{T_j} \sim N(0,\sigma_{T_j}^2 = 10^{-6})$.

\begin{figure}
	\vspace{-1em}
	\centering
	\subfloat[]{\includegraphics[width=0.8\linewidth]{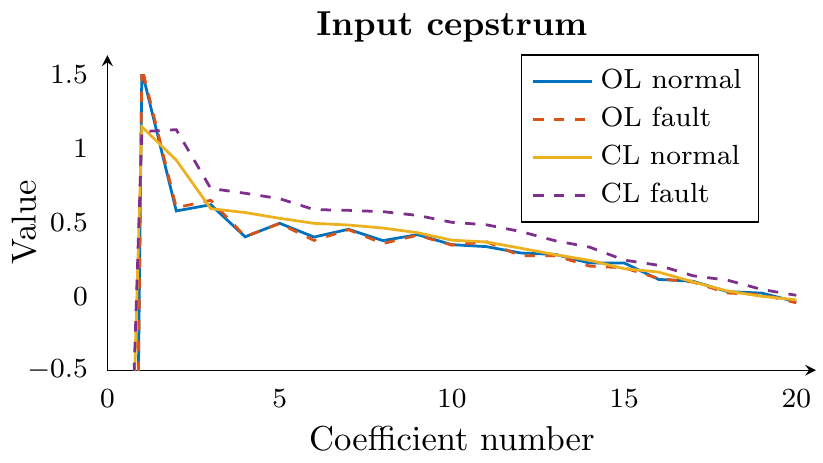}
	\label{fig:input}}

	\subfloat[]{\includegraphics[width=0.8\linewidth]{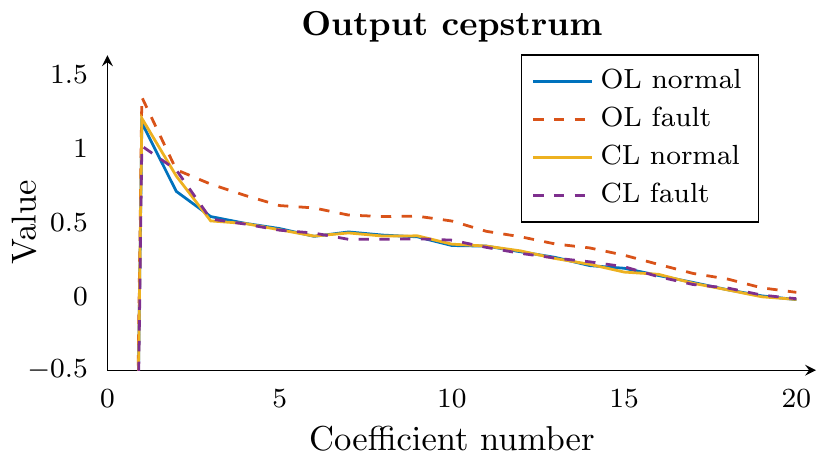}
	\label{fig:output}}                                
	
	\subfloat[]{\includegraphics[width=0.8\linewidth]{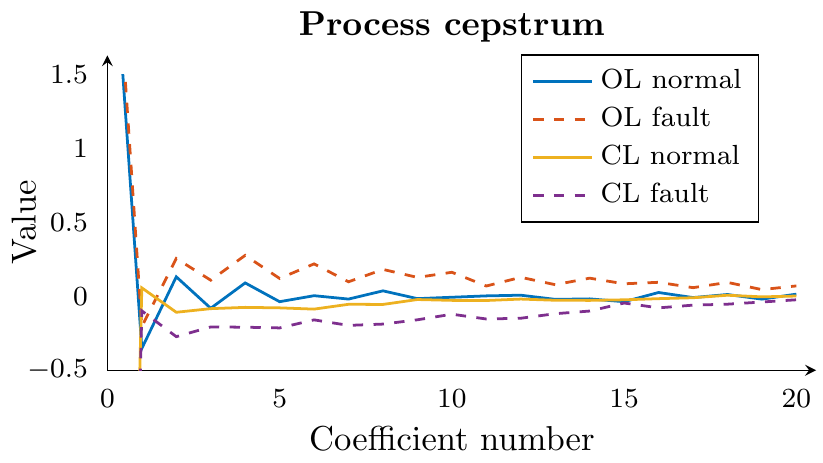}
	\label{fig:process}}    

\label{fig:cepstra}   
\caption{The cepstrum coefficients for input, output and process. We show both normal and faulty operating behaviour, with (CL) and without (OL) the controller. The cepstral coefficients for the process in (c) clearly show different fingerprints for the different cases. The normal operating behaviour changes only slightly when turning on the controller, capturing the extra dynamics of the controller. However, faulty operating behaviour results in a deviation from the normal operating behaviour, both in OL and, notably, in CL, capturing faults hidden by the controller. We can see the different contributions to this process cepstrum (see Equation \eqref{eq:additive}) in (a) and (b): (a) shows the change in the input dynamics in the faulty regime, when the controller starts compensating. (b) shows the change in the output in the faulty regime when the controller is turned off.}                         
\end{figure}

We estimate the cepstra of input, output and of the process itself and show that the MIMO cepstrum indeed captures the dynamics of the processes and controller involved. The 200 data points around the fault are omitted.
Results are shown and discussed in Fig. \ref{fig:input} for the input signals, Fig. \ref{fig:output} for the output signals and Fig. \ref{fig:process} for the process dynamics. Notice that the cepstrum of the process explicitly shows the change in the reactor dynamics, whether the controller is on or off, and detects hidden faults in a process, without having to model the process.

This makes the MIMO cepstrum a very promising technique for building new anomaly detection or fault prognosis algorithms, for example by extending and employing the distance measure from \cite{DeCockThesis,lauwers2017,martin2000metric} in a clustering algorithm.

\section{CONCLUSIONS AND FUTURE WORK}
\label{sec:conclusions}
In this paper, we introduced a new definition for the power cepstrum, that extends it to systems with multiple inputs and outputs, based on the Smith-McMillan form of the system. This new power cepstrum is then interpreted in terms of the poles and zeros of the underlying models. For systems with as many inputs as outputs, we provide a method to calculate the power cepstrum based on input and output data.

We then illustrate the theoretical results from this paper with a numerical example on a simple synthetic model, and on a realistic dataset concerning a Non-isothermal Continuous Stirred Tank Reactor. We show that the methods and interpretations presented here indeed hold numerically.

The Reactor case study shows potential to leverage the extended power cepstrum as an anomaly detection technique. We believe that every fault will have its own "fingerprint" in the cepstrum domain, which will then allow us to discern between different (perhaps compounded) types of faults.

Future work includes extending the distance in \cite{lauwers2017} to the MIMO case, providing data-driven ways to compute the cepstrum in the case of unequal number of inputs and outputs and proving links with canonical correlations and mutual information, as in the SISO case \cite{DeCockThesis}.

\addtolength{\textheight}{-3cm}   



\section*{APPENDIX}

In this Appendix, we derive the results in Equation \eqref{eq:cepstrumpoleszeros}.
Starting from Equation \eqref{eq:transferfunctionpoleszeros} and using Equation \eqref{eq:powerspectrum}, we write
\begin{equation}
\begin{aligned}
&\log(\Phi_H(\textnormal{e}^{i\omega})) = \log (H(\textnormal{e}^{i\omega})\overline{H(\textnormal{e}^{i\omega})})\\
&\hspace{-1em}= \log(g^2) + \sum_{i=1}^{q}\left(\log\left(1 - \beta_i \textnormal{e}^{-i\omega}\right) + \log\left(1 - \overline{\beta}_i \textnormal{e}^{i\omega}\right)\right)\\
&- \sum_{i=1}^{p}\left(\log\left(1 - \alpha_i \textnormal{e}^{-i\omega}\right) + \log\left(1 - \overline{\alpha}_i \textnormal{e}^{i\omega}\right)\right).
\end{aligned}
\end{equation}
Employing the series expansion
\begin{equation}
\log(1-x) = -\sum_{k = 1}^{\infty}\frac{x^k}{k} \hspace{7pt} \forall |x|<1,
\label{eq:logseries}
\end{equation}
we find
\begin{equation}
\begin{aligned}
\log(\Phi_H(&\textnormal{e}^{i\omega}))\\ 
\hspace{-0.1em} = \log(g^2) & - \sum_{i=1}^{q}\left(\sum_{k=1}^\infty \frac{\beta_i^k}{k}\textnormal{e}^{-ik\omega} + \sum_{k=1}^\infty \frac{\overline{\beta}_i^k}{k}\textnormal{e}^{ik\omega}\right)\\
&+ \sum_{i=1}^{p}\left(\sum_{k=1}^\infty \frac{\alpha_i^k}{k}\textnormal{e}^{-ik\omega} + \sum_{k=1}^\infty \frac{\overline{\alpha}_i^k}{k}\textnormal{e}^{ik\omega}\right).
\end{aligned}
\label{eq:logspectrum}
\end{equation}
The final step consists of noting that the $c_H(k)$, are the inverse Fourier transform of $\log(\Phi_H)$, or
\begin{equation}
\sum_{k=-\infty}^{\infty}c_H(k)e^{-ik\theta} = \log\Phi_H(e^{i\theta}).
\label{eq:inversefouriertransform}
\end{equation}
Matching Equations \eqref{eq:inversefouriertransform} and \eqref{eq:logspectrum}, we obtain Equation \eqref{eq:cepstrumpoleszeros}.


\bibliographystyle{IEEEtran}

\bibliography{references}

\begin{thebibliography}{10}
\providecommand{\url}[1]{#1}
\csname url@samestyle\endcsname
\providecommand{\newblock}{\relax}
\providecommand{\bibinfo}[2]{#2}
\providecommand{\BIBentrySTDinterwordspacing}{\spaceskip=0pt\relax}
\providecommand{\BIBentryALTinterwordstretchfactor}{4}
\providecommand{\BIBentryALTinterwordspacing}{\spaceskip=\fontdimen2\font plus
\BIBentryALTinterwordstretchfactor\fontdimen3\font minus
  \fontdimen4\font\relax}
\providecommand{\BIBforeignlanguage}[2]{{%
\expandafter\ifx\csname l@#1\endcsname\relax
\typeout{** WARNING: IEEEtran.bst: No hyphenation pattern has been}%
\typeout{** loaded for the language `#1'. Using the pattern for}%
\typeout{** the default language instead.}%
\else
\language=\csname l@#1\endcsname
\fi
#2}}
\providecommand{\BIBdecl}{\relax}
\BIBdecl

\bibitem{bogert1963quefrency}
B.~P. Bogert, M.~J. Healy, and J.~W. Tukey, ``The quefrency alanysis of time
  series for echoes: Cepstrum, pseudo-autocovariance, cross-cepstrum and saphe
  cracking,'' in \emph{Proceedings of the symposium on time series analysis},
  vol.~15, 1963, pp. 209--243.

\bibitem{oppenheim1975digital}
A.~Oppenheim and R.~Schafer, \emph{Digital Signal Processing (First
  edition)}.\hskip 1em plus 0.5em minus 0.4em\relax Prentice-Hall Englewood
  Cliffs, NJ, 1975.

\bibitem{randall2017history}
R.~B. Randall, ``A history of cepstrum analysis and its application to
  mechanical problems,'' \emph{Mechanical Systems and Signal Processing},
  vol.~97, pp. 3--19, 2017.

\bibitem{liu2016extreme}
H.~Liu, L.~Yu, W.~Wang, and F.~Sun, ``Extreme learning machine for time
  sequence classification,'' \emph{Neurocomputing}, vol. 174, pp. 322--330,
  2016.

\bibitem{DeCockThesis}
K.~De~Cock, ``Principal angles in system theory, information theory and signal
  processing,'' \textit{PhD thesis}, 293 pp, KU Leuven, May 2002.

\bibitem{lauwers2017}
O.~Lauwers and B.~De~Moor, ``A time series distance measure for efficient
  clustering of input/output signals by their underlying dynamics,'' \emph{IEEE
  Control Systems Letters}, vol.~1, pp. 286--291, June 2017.

\bibitem{kailath1980linear}
T.~Kailath, \emph{Linear systems}.\hskip 1em plus 0.5em minus 0.4em\relax
  Prentice-Hall Englewood Cliffs, NJ, 1980.

\bibitem{welch1967use}
P.~Welch, ``The use of fast fourier transform for the estimation of power
  spectra: a method based on time averaging over short, modified
  periodograms,'' \emph{IEEE Transactions on audio and electroacoustics},
  vol.~15, no.~2, pp. 70--73, 1967.

\bibitem{de2000subspace2}
K.~De~Cock and B.~De~Moor, ``Subspace angles and distances between arma
  models,'' in \emph{Proc. of the Intl. Symp. of Math. Theory of networks and
  systems}, vol.~1, 2000.

\bibitem{knill2014cauchy}
O.~Knill, ``Cauchy--binet for pseudo-determinants,'' \emph{Linear Algebra and
  its Applications}, vol. 459, pp. 522--547, 2014.

\bibitem{vishwanathan2004binet}
S.~Vishwanathan and A.~J. Smola, ``Binet-cauchy kernels,'' in \emph{Proceedings
  of the 17th International Conference on Neural Information Processing
  Systems}.\hskip 1em plus 0.5em minus 0.4em\relax MIT Press, 2004, pp.
  1441--1448.

\bibitem{martin2000metric}
R.~J. Martin, ``A metric for arma processes,'' \emph{Signal Processing, IEEE
  Transactions on}, vol.~48, no.~4, pp. 1164--1170, 2000.

\bibitem{cstr_paper}
G.~Li, S.~J. Qin, Y.~Ji, and D.~Zhou, ``Reconstruction based fault prognosis
  for continuous processes,'' \emph{Control Engineering Practice}, vol.~18,
  no.~10, pp. 1211 -- 1219, 2010.

\end{thebibliography}

\end{document}